\documentclass[10pt,twocolumn,letterpaper]{article}

\usepackage[spanish,english]{babel}
\usepackage[utf8x]{inputenc}
\usepackage[T1]{fontenc}
\usepackage{array,tabularx}
\usepackage{algorithm2e}
\usepackage{verbatim}
\usepackage{tikz}

\usepackage[a4paper,top=3cm,bottom=2cm,left=3cm,right=3cm,marginparwidth=1.75cm]{geometry}

\usepackage{amsmath}
\usepackage{graphicx}
\usepackage[colorinlistoftodos]{todonotes}
\usepackage[colorlinks=true, allcolors=blue]{hyperref}
\usepackage{subcaption}

\newenvironment{conditions*}
  {\par\vspace{\abovedisplayskip}\noindent
   \tabularx{\columnwidth}{>{$}l<{$} @{\ : } >{\raggedright\arraybackslash}X}}
  {\endtabularx\par\vspace{\belowdisplayskip}}

\title{
		\usefont{OT1}{bch}{b}{n}
		\normalfont \normalsize \textsc{} \\ [10pt]
		\huge Player Availability Rating (PAR) - A Tool for Quantifying Skater Performance for NHL General Managers  \\
}

\usepackage{authblk}

\author[1]{Shuja Khalid}

	\affil[1]{\small{Department of Computer Science, University of Toronto}}

\begin{document}
\maketitle

\selectlanguage{english}
\begin{abstract}
This project aims to assess the performance of various regression models in predicting the performance of hockey players. The measure of performance is chosen to be points scored (sum of goals scored and assists made) by individual players per game (PPG). This paper thus considers the offensive performance of the players and uses PPG as a metric to assign a value to the player. This predicted value can be used to rank players and is similar to what TSN.com\cite{tsn} and NHL.com\cite{nhl} use to rank players before each hockey season. A combination of physical characteristics, shooting percentage, and usage during games are used as training features. A novel metric, Player Availability Rating (PAR) is proposed which utilizes the offensive predictions to predict the availability of a player during a season. These metrics can be valuable in the NHL as it would allow for general managers to track a players performance and their availability during the course of a season.
\end{abstract}


\section{Introduction}
The core competency required of a manager is to be able to assess a players performance. This is a challenging endeavor as the performance of a player depends on a number of attributes and understandably varies from season to season. The approach adopted by this paper is to create a model that uses only player characteristics such as height, weight, expected time on ice and expected shooting percentage to create a projection for a player. Various regression based models were created in order to estimate the offensive production of a players in the form of points per game (PPG). Neural Nets were found to produce the estimates that were closest to the established baseline. This gives managers and coaches a way to quantify a player’s contribution to the team based on a model that was trained using data accumulated from 2006-2017. This model can also serve as a guide to how players entering the league, for whom previous National Hockey League data is not available, will perform based on their physical attributes and their usage during an 82 game NHL season. The paper takes these projections a step further and proposes a metric that uses the projected player contributions (PPG), their actual performance during the current season (PPG), and the team’s performance (points percentage) to rank players that might be available for trade. This metric can thus serve as a useful tool for any general manager, real or fantasy to make roster decisions.
\section{Process Diagram}
\begin{figure}[h!]
  \centering
  \includegraphics[scale=0.35]{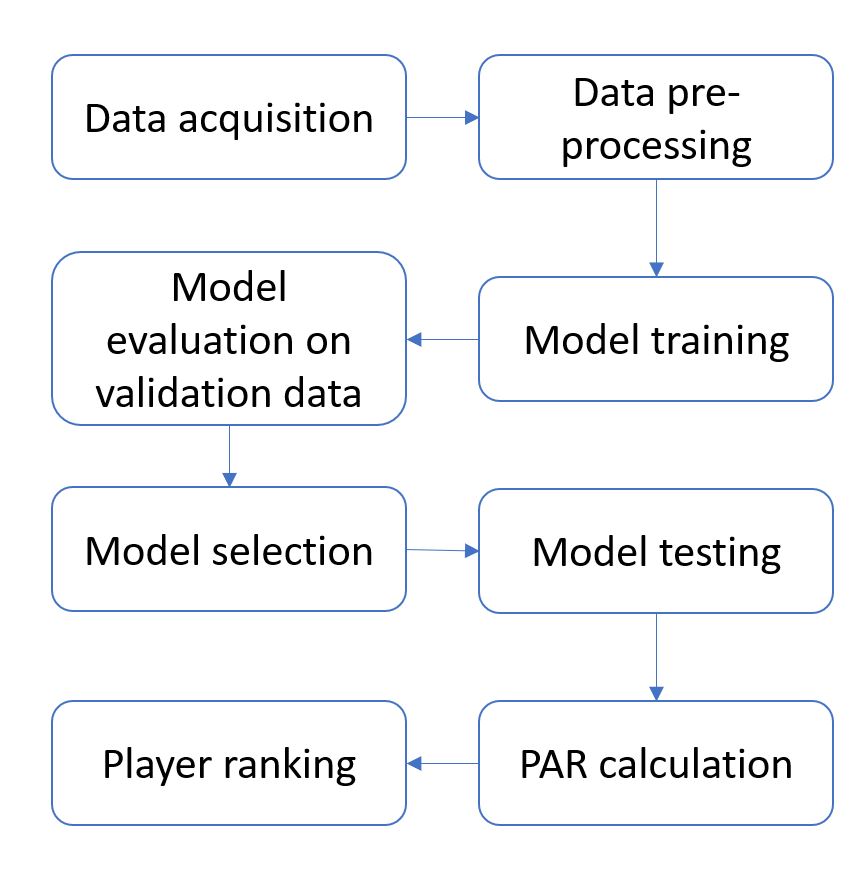}
  \caption{Visual representation of flow of data for producing the PAR rating}
  \label{fig:flowchart}
\end{figure}

\section{Formal Description}

The performance of players in the National Hockey League varies from season to season due to a variety of reasons such as overall team performance, player usage, physiological attributes, coaching styles etc. Each of the 31 teams in the league employ an army of scouts that are responsible for analyzing players over the course of a season which can be a daunting task. This paper proposes a tool that can be used by general managers to evaluate the performance of players during the course of a season based on their expected and predicted offensive performances, the long-term and short-term performances of their teams, and their usage per game. The algorithm used to determine the performance of the players is presented in Algorithm 1.

\begin{algorithm}
\SetAlgoLined
\KwResult{ PAR (Player Availability Rating) }
 $PPG_{predicted}$, $PPG_{actual}$, $PPCG_{season}$, $PPCG_{recent}$, $X_{test}$, $Y_{test}$, $X_{train}$, $Y_{train}$, $w_o=2$\;
 method=\{'linearRegression', 'k-NN', 'NeuralNets', 'decisionTree', 'randomForest'\}\;
 $model_{minerror}$ = Choose method that produces the lowest average error\;
 \While {items in $X_{test}$}{
  Use $model_{minerror}$ with lowest calculated error and recalculate $PPG_{predicted}$\;
  Extract latest PPG values from NHL.com and assign to $PPG_{actual}$\;
  Extract team PPCG values (season) from NHL.com and assign to $PPCG_{season}$\;
  Extract team PPCG values (recent) from NHL.com and assign to $PPCG_{recent}$\;
  PAR = $\frac{(PPG_{predicted}-PPG_{actual})}{(PPCG_{season})} + w_o*\frac{(PPG_{predicted}-PPG_{actual})}{(PPCG_{recent})}$\;
 }
 \caption{PAR Algorithm}
\end{algorithm}

Each of the models were implemented by using existing libraries (scikit-learn) and were optimized by running a grid search over the parameters. 

This insight is invaluable as the players on the list above are either severely under-performing, or their teams are not performing well or a combination of both. These players may be more likely to be surrendered by opposing general managers during trade negotiations and might be considered under the radar acquisitions with the potential for very high reward. The formula used to determine this rating is presented below:
\\
\\
$
PAR = \frac{(PPG_{predicted}-PPG_{actual})}{(PPCG_{season})} 
	+ w_o*\frac{(PPG_{predicted}-PPG_{actual})}				{(PPCG_{recent})}
$
\\
\\
where
\begin{conditions*}
PPG_{predicted} & Predicted points/game for skater \\
PPG_{actual} & Actual points/game for skater \\
PPCG_{season} & Percentage of points earned by team during current season \\
PPCG_{recent} & Percentage of points earned by team during last 10 games \\
w_{o} & A parameter to control the influence of recent team performance on player availability \\
\end{conditions*}

The difference between the predicted and actual PPG values for each player is indicative of the player’s performance. A negative value corresponds to the player out-performing expectations whereas a positive value means that the player is under-performing. The short term and long term winning percentages of the player’s team can be indicative of the pressure that the opposing general manager might be facing to make a move to either trade for a player or to trade an under-performing player. Based on other factors such as health and confidence levels, which are challenging to quantify, the player might simply need a change of scenery and a trade to a different team might be a winning proposition for all parties involved. Regularly using the proposed algorithm can help managers stay on top of such situations.

The PAR estimate captures the essence of existing ratings that are dependent on probabilistic considerations. However, the formulation presented above is not derived from any of these sources. It also does not use a probabilistic method to make predictions. 

The PAR estimate uses a combination of neural nets and empirical formulation to quantify the availability of an individual during an NHL season. Such a formulation was not found in literature. However, a number of sources attempted to quantify player and team performance based on a probabilistic approach \cite{trueskill}. The most popular one being TrueSkill \cite{trueskill} which is a Bayesian Rating system that has been applied to other sports such as Basketball \cite{ncaa} and Football \cite{ncaa}.  

\section{Related Work}
The following sources were consulted during literature review:
\newline

i.) \textit{Forecasting Success in the National Hockey League using In-Game Statistics and Textual Data \cite{predict}:} 

This paper utilizes traditional and advanced statistics for individual players on a team to predict how teams will perform over the course of a season. The core concept of using statistics to determine the cumulative performance of players is similar to the idea presented in this paper. However, PAR makes a point not to use advanced statistics and in its stead makes use of physical player characteristics and their expected usage over the course of an NHL season.
\newline

ii.) \textit{Estimating the Value of Major League Baseball Players \cite{fields}:} 
This paper attempts to quantify the value of players to determine how much they should be paid. The author proposes a formulation that considers a number of features/factors that might determine player value. PAR attempts to consider similar features but only looks at offensive contribution.
\newline

iii.) \textit{Predicting the Major League Baseball Season \cite{jia}:} 
This paper uses neural networks to solve a binary classification problem in the form of wins and losses for baseball teams over the course of a Major League Baseball season. Their use of neural networks along with a large amount of data to make these predictions.
\newline

iv.) \textit{TrueSkill - A Bayesian skill rating system \cite{trueskill}:} 
The paper above uses a probabilistic approach to skill assessment to produce a rating based on the outcome of previously played games. This paper uses chess rankings to illustrate their approach.
\newline

v.) \textit{Knowing what we don't know in NCAA Football ratings: Understanding and using structured uncertainty \cite{ncaa}:} 
This paper uses the TrueSkill method and applies it to evaluate team performance for NCAA football games. The focus is on team performance as opposed to player performance.   
\newline

The papers reviewed above focus on a probabilistic evaluation of performance. After an exhaustive search of the literature, no papers were found that use non-probabilistic machine learning algorithms to produce real valued outputs to evaluate player contributions (offensive or defensive). An attempt had been made to do so in baseball but it was limited to the outcome of games based on recent performance \cite{trueskill}.

\section*{Comparison or Demonstration}
In order to demonstrate the effectiveness of neural nets to predict player performance based on historic data, five models were created using five different regression based methods. The performance of the neural network was compared against these methods. The table below summarizes the error values observed (predicted PPG-actual PPG) for these methods. Table X also looks at predictions made by TSN.ca and NHL.com before the start of the 2017-2018 hockey season and compares them to the baseline (current PPG values) for the top 100 players in the league as determined by their PPG.

The training, validation and test datasets were created by creating scripts that extracted the required data from NHL.com and using an 80-10-10 split. Similarly, additional scripts were created in order to extract baseline data from NHL.com\cite{nhl} and TSN.ca\cite{tsn}. Each feature was modified using z-score normalization before training the model.

Based on the results in Table 1, neural nets provided the lowest error values of any method with linear regression having the worst performance of any method, as expected. Figure 2 illustrates the performances of each of the methods in predicting the top 100 players with the highest PPG values. 

\begin{table}
\centering
\label{my-label}
\begin{tabular}{|l|l|l|}
\hline
Method/Source     & Mean & Median \\ \hline
Neural Nets       & 0.211      & 0.188        \\ \hline
Decision Tree     & 0.222      & 0.21         \\ \hline
Random Forest     & 0.215      & 0.193        \\ \hline
k-NN Regression   & 0.234      & 0.21         \\ \hline
Linear Regression & 0.245      & 0.22         \\ \hline
TSN.ca            & 0.202      & 0.173        \\ \hline
NHL.com           & 0.197      & 0.167        \\ \hline
\end{tabular}
\caption{Calculated error values comparing predicted and actual PPG values for top 100 scorers in the NHL}
\end{table}

\begin{table}
\centering
\label{my-label}
\begin{tabular}{|c|c|c|}
\hline
Payer Name       & Actual & Predicted \\ \hline
Nikita Kucherov  & 1.48    & 1.18      \\ \hline
Brad Marchand    & 1.17    & 1.02      \\ \hline
Claude Giroux    & 1.07    & 1         \\ \hline
Connor McDavid   & 1.17    & 0.99      \\ \hline
Anze Kopitar     & 1.17    & 0.99      \\ \hline
Johnny Gaudreau  & 1.28    & 0.98      \\ \hline
John Tavares     & 1.14    & 0.97      \\ \hline
Evgeny Kuznetsov & 1.06    & 0.96      \\ \hline
Brayden Point    & 0.85    & 0.93      \\ \hline
Mark Scheifele   & 1.21    & 0.91      \\ \hline
\end{tabular}
\caption{Comparison of actual and predicted PPG values for the current top 10 offensive contributors in the NHL}
\end{table}

\begin{figure}[h!]
  \centering
  \begin{subfigure}[b]{0.4\linewidth}
    \includegraphics[width=\linewidth]{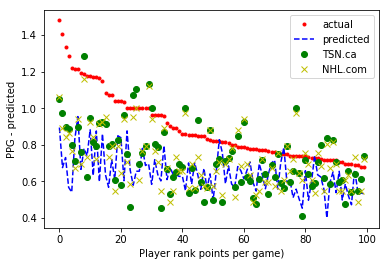}
     \caption{Linear Regression}
  \end{subfigure}
  \begin{subfigure}[b]{0.4\linewidth}
    \includegraphics[width=\linewidth]{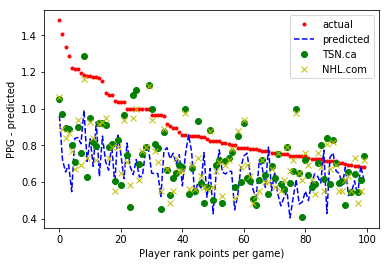}
    \caption{k-Nearest Neighbors}
  \end{subfigure}
  \begin{subfigure}[b]{0.4\linewidth}
    \includegraphics[width=\linewidth]{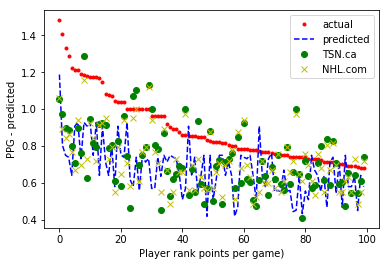}
    \caption{Decision Trees}
  \end{subfigure}
  \begin{subfigure}[b]{0.4\linewidth}
    \includegraphics[width=\linewidth]{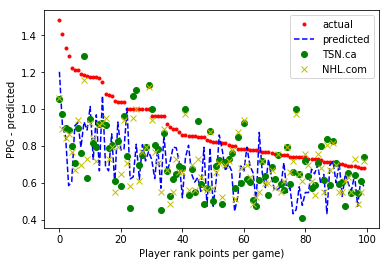}
    \caption{Random Forest}
  \end{subfigure}
  \begin{subfigure}[b]{1.0\linewidth}
    \includegraphics[width=\linewidth]{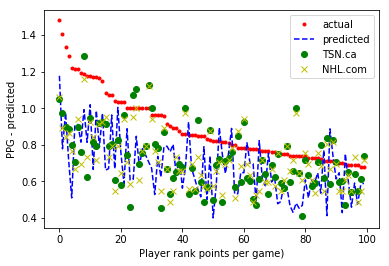}
    \caption{Neural Networks}
  \end{subfigure}
  \caption{Comparison of various regression methods against baselines for the top 100 players with the highest PPG values in the NHL}
  \label{fig:coffee3}
\end{figure}

The top 10 players based on the predicted PPG determined by the neural nets is presented in Table 2. Their current PPG values are also presented as a reference. The difference in their predicted vs actual values can be attributed to the current season being only 30\% complete. Over the course of the season, the actual PPG values are expected to decrease.  

The final step of the PAR algorithm is to apply the PAR formula using the predicted PPG projections, baseline data and team-based statistics. The top 10 players with the highest PAR values are presented in Table 3.

\begin{table}
\centering
\label{my-label}
\begin{tabular}{|c|c|}
\hline
Player Name       & PAR  \\ \hline
Ryan Dzingel      & 2.29 \\ \hline
Mark Stone        & 2.07 \\ \hline
Cam Fowler        & 2.03 \\ \hline
Brandon Montour   & 1.66 \\ \hline
Tyler Myers       & 1.64 \\ \hline
Brendan Perlini   & 1.21 \\ \hline
Nick Foligno      & 1.14 \\ \hline
Tomas Tatar       & 1.12 \\ \hline
Dion Phaneuf      & 1.02 \\ \hline
Gabriel Landeskog & 0.98 \\ \hline
\end{tabular}
\caption{Top 10 players with the highest PAR estimates}
\end{table}

\section*{Limitations}
The presented model provides estimates that are far from the norm for players that are highly skilled. An example of this behavior is for players that are very large in size (taller than 6’ 3’’ and weigh more than 220 pounds). Over the course of the last 10 years,  the majority of such players have traditionally been enforcers (players that are not relied upon to score points). Only a handful of players that meet this criteria are prolific scorers. The model thus assigns low point per game estimates to these types of players. An example of this is Patrik Laine who was assigned a value of 0.63. This was extremely surprising because his actual performance in his first year in the league far exceeded the estimate (by a factor of 1.5). Similarly, players that are smaller in size (smaller than 5’ 9’’ and weigh less than 170 pounds) are more likely to be assigned much higher point per game values because traditionally such players have scored at very high rates.

One of the most important ways to improve this model is to include more features. These features should be a combination of player characteristics and situational usage. Another possible extension of this project would be to assign monetary values to these players in an attempt to present what their valuations should be at any point during the season. The incorporation of the methodology in the paper by Fields \cite{predict} would draw upon various features related to team performance, situational usage and situational performance to present estimates of valuations for players. Such a tool would be invaluable for general managers in the National Hockey League as they attempt to assemble their rosters while working under tight constraints such as the unavailability of funds (for managers of small market teams), and a salary cap enforced by the National Hockey League which limits the amount of money that each team can spend on players.  

The prediction criteria defined in the algorithm makes it unique as it can also be applied to other sports such as Basketball, Baseball and Football. This is another area that might be worth exploring in the future.

\section*{Conclusions}
The goal of the paper was to assess the viability of using neural nets to predict player performance. The results indicate that neural nets and other regression based methods can be used to adequately complete this task. The results were compared to actual PPG values as well as other sources such as TSN.ca\cite{tsn} and NHL.com\cite{nhl} that are considered to be a top resource for player projections. The project was also extended in order to predict the estimate the Player Availability Rating (PAR) which is a novel metric that aims to quantify the availability of a player based on the current performance of the player and the performance of their team. The results indicate that neural nets outperform the other regression based methods and are comparable to those made by TSN.ca\cite{tsn} and NHL.com\cite{nhl}.

The author has also launched a website that has adopted the  algorithm presented in this paper:
\href{https://gmaiplaybook.com}{gmaiplaybook.com}
\\

\bibliographystyle{ieee} 
\bibliography{references.bib}

\end{document}